\documentclass{article}                    
\usepackage{graphicx}
\usepackage{latexsym}
\newcommand{\be}{\begin{equation}}
\newcommand{\ee}{\end{equation}}
\newcommand{\bea}{\begin{eqnarray}}
\newcommand{\eea}{\end{eqnarray}}

\begin{document}

\title{Casimir Effect and Free Fall\\
 in a Schwarzschild Black Hole}

\author{Francesco Sorge\footnote{e-mail: sorge@na.infn.it}\\        
I.N.F.N.\, Naples\\
Naples, I-80126, Italy\\
}

\maketitle

\begin{abstract}
Scalar field vacuum energy and particle creation in a 3D Casimir apparatus, freely falling in the Schwarzschild spacetime, are considered in the reference frame of a comoving observer. Following Schwinger's proper time approach,  Casimir energy is evaluated from the effective action, resulting in a small correction to the flat spacetime case. Besides, a tiny amount of quanta excited out from the vacuum is found. Both effects are discussed, drawing attention to the role of the underlying spacetime dimensionality.

[{\em Talk given at the 15$^{th}$ Marcel Grossmann Meeting,  1- 7 July, 2018, Rome, Italy}]
\end{abstract}
 \vspace{4pt}
 
{\bf keywords}: Casimir effect; quantum fields in curved space-times; particle creation.

%%%%%%%%%%%%%%%%% now a standard article style for the most part

\section{Introduction}
Roughly speaking, Casimir effect \cite{casimir1,casimir2,casimir3} originates from a distortion in the modes of a quantum field constrained in a finite region of space by some boundaries. The latter can be material as well as due to the geometrical properties of the background spacetime \cite{calloni1,calloni2,marquez3,marquez4,sorge1,fulling1,milton,sorge2,sorge3}. 

Here we will show that an observer comoving with a Casimir cavity, freely falling in a Schwarzschild black hole, measures a small reduction in the (absolute) value of the (negative) Casimir energy  as the black hole horizon is approached.  At a first glance, this may seem rather puzzling, as one would expect no change with respect to the usual flat spacetime result $\langle\epsilon_{Cas}\rangle_{stat}=-\frac{\pi^2}{1440L^4}$, due to the Equivalence Principle. Actually, the {\em local} measurements performed by the comoving observer are related to {\em non-local} quantities, as the effective action $W$ and the renormalized stress-energy tensor. The latter are determined (through various  regularization techniques) by the low-energy, long-wavelength contributions of the field modes, thus probing the {\em global} structure of the surrounding spacetime geometry.  
This, in turn, allows for a local measurement to be sensitive to the cavity fall.

When studying the influence of a gravito-inertial environment on the Casimir energy, we may consider
\begin{itemize}
\item{{\em tidal} effects: due to the spatial extension of the Casimir apparatus, they are expected to cause anisotropies in the distribution of the vacuum energy density inside the cavity. Such effects will be discussed in \cite{justin}, by means of a detailed analysis of 1+1D model.}
\item{{\em non-local} effects: also when tidal effects are neglected, we can still face possible corrections to Casimir energy, closely related to the non-locality of the quantum field theory obeyed by the field confined in the cavity.  In spite of the geodesic {\em inertial} free fall, the {\em non-local} nature of the field stress-energy tensor allows for both the global properties of the background spacetime {\em and} the cavity motion to influence the {\em local} measurements performed by an observer comoving with the cavity.}
\end{itemize}

In this paper we will focus on this latter aspect. 

In section 2 we introduce the Lema\^itre coordinates. In section 3 we give some basic assumptions. Section 4 is devoted to Schwinger's proper-time approach and to the evaluation of the effective action $W$. Section 5 deals with the static Casimir effect, while particle creation is discussed in section 6. The last section is devoted to some concluding remarks.
Throughout the paper use has been made of natural geometrized units. Greek indices take values from 0 to 3; latin ones take values from 1 to 3. The metric signature is $-2$, with determinant $g$.

\section{Lema\^itre Coordinates}
The metric for a black hole of mass $M$ in the standard Schwarzschild coordinates $\{t,r,\theta,\phi\}$ reads
\be\label{schw}
ds^2= \bigg(1-\frac{r_g}{r}\bigg)dt^2-\bigg(1-\frac{r_g}{r}\bigg)^{-1}dr^2-r^2 d\Omega^2,
\ee
where $r_g=2M$ is the gravitational (Schwarzschild) radius of the black hole and $d\Omega^2=d\theta^2+\sin^2\theta d\phi^2$.  For the present analysis, we need a coordinate set which is regular at the horizon. We will adopt the Lema\^itre  coordinates $\{\tau, \rho, \theta, \phi\}$, in which a freely falling test body has a constant value $\rho_0$ of the radial $\rho$ coordinate.

Consider a Casimir cavity, freely falling from spatial infinity. Adjust the comoving observer clock so that the proper time $\tau=0$ when the cavity is at the radial horizon coordinate $r_0=r_g$. Then $\rho_0=\frac{2}{3}r_g$ and the Schwarzschild metric in the Lema\^itre coordinates reads \cite{lemaitre,kramer} 
\be\label{lm}
ds^2=d\tau^2-\frac{r_g}{r(\tau)}d\rho^2-r^2(\tau) d\Omega^2,\quad\quad\quad r(\tau)=r_g\bigg(1-\frac{3\tau}{2r_g}\bigg)^{2/3}.
\ee
Notice that the travel from the infinity to the horizon is described by {\em negative} values of the proper time: $-\infty<\tau\leq 0$. 

\section{Basic Assumptions}
Some basic assumptions are in order. The measurement of the Casimir energy inside the falling cavity is performed by a {\em comoving}  observer.  The cavity plates (of area $A$ and separated by a distance $L$, such that $L\ll\sqrt{A}$) are taken orthogonal to the radial falling direction\footnote{Such a choice has been made only for the sake of definiteness.}. Furthermore
\begin{itemize}
\item{the cavity is falling from spatial infinity with zero initial velocity and zero angular momentum;}
\item{the cavity size is much smaller than the gravitational radius of the black hole, so that, in particular, $L\ll r_g$, with $L$ being the  plate separation;}
\item{the cavity is {\em rigid}; its dimensions and shape do not suffer any distortion. Such assumption holds true  provided $L\ll r_g$. Differently stated, we neglect tidal effects. Tidal effects will be extensively studied in a lower-dimensional model in \cite{justin};}
\item{the cavity follows a true geodesic motion; other non-gravitational external effects are neglected.} 
\end{itemize}

\section{Schwinger Proper-Time Approach}

For the sake of simplicity we will consider a massless scalar field. We also assume the field to obey the Dirichlet boundary conditions at the plates. From the diagonal metric (\ref{lm}) the tetrad adapted to the comoving observer is readily obtained, as well as the Klein-Gordon equation for the (minimally coupled) scalar field \cite{birrell}
\be\label{eqdiff1}
\bigg[\Box +\frac{1}{4}\frac{\xi^2}{(1-\xi\tau)^2}\bigg]\varphi=0,\quad\quad\quad \xi=\frac{3}{2r_g}.
\ee
We guess the following solution
\be\label{phi}
\varphi(x^a)\sim e^{i\vec k_\perp\cdot \vec x_\perp}\sin\bigg(\frac{n\pi}{L}x\bigg)\chi(\tau),\quad\quad\quad\quad n\in N
\ee
where $\vec k_\perp\equiv (k_y, k_z)$ and $\vec x_\perp\equiv (y, z)$. We point out that the same spatial dependence as in the flat spacetime case has been assumed {\em a priori}, since we have supposed no relevant tidal effects inside the cavity. $\chi(\tau)$ is a function of the proper (local) time, to be evaluated below.
Plugging (\ref{phi}) in (\ref{eqdiff1}) we get the following equation for $\chi(\tau)$
\be\label{chiequation}
\bigg[\partial_\tau^2 +\omega_k^2 +\frac{1}{4}\frac{\xi^2}{(1-\xi\tau)^2}\bigg]\chi=0,
\ee
where $\vec k\equiv(n\pi/L, \vec k_\perp)$ and $\omega_k^2=k^2_\perp+\big(n\pi/L\big)^2$. 
Finally, the field modes $\chi_k(\tau)$ are found in terms of a zero-order Hankel function $H_0^{(1)}$
\be\label{chitot}
\chi_k(\tau)=\frac{1}{2}\sqrt{\frac{\pi}{\xi}(1-\xi\tau)}H^{(1)}_0\bigg(\frac{\omega_k}{\xi}(1-\xi\tau)\bigg).
\ee

From (\ref{eqdiff1}), the proper-time Hamiltonian $\hat H$  reads $\hat H = \hat H_0+\hat V$, where $\hat H_0= \partial_\tau^2-\vec\nabla^2\equiv -\hat p_0^2+\hat{\vec p}^2$.
As usual, we write the effective action $W=\lim_{\nu\rightarrow 0}W(\nu)$,  where \cite{schwinger1,schwinger2}
\be\label{W}
W(\nu)=-\frac{i}{2}\int_0^\infty\, ds\, s^{\nu-1}{\rm Tr}\, e^{-is\hat H},
\ee
and the limit $\nu\rightarrow 0$ has to be taken at the end of calculations. The trace has to be evaluated all over the continuous as well the discrete degrees of freedom, including those of spacetime. After a quite long algebra (involving a lot of special function properties) we get \cite{grad,nist}
%\begin{widetext}
\bea\label{totalW}
W(\nu)=-\frac{iA}{32\pi^{5/2}}\int_0^\infty ds \int_{-\infty}^Td\tau\sum_n \frac{s^{\nu-3/2-1}}{\beta^{1/2}}  e^{-is(n\pi/L)^2}\nonumber\\
\times \bigg[\pi^{3/2}e^{-i/(2\beta)}H_0^{(1)}\big(1/(2\beta)\big)+2G_{23}^{31}\bigg(-\frac{i}{\beta}
\left |
\begin{array}{cc}
 0 &   1/2  \\
0&0\,\,\,0
\end{array}
\right)\bigg],
\nonumber
\\
\eea
%\end{widetext}
where $\beta=\frac{s\xi^2}{(1-\xi\tau)^2}$ is a small adimensional parameter (which will be used for further power expansion) and $G_{23}^{31}$ is a Meijer G-function. Inspection of (\ref{totalW}) shows that the $H_0^{(1)}$-dependent part of $W(\nu)$ is {\em real}, hence  responsible for vacuum polarization (namely, the static Casimir effect), while the $G$-dependent part of $W(\nu)$ is {\em imaginary} and related to the vacuum persistence amplitude (i.e., particle creation).

\section{The Static Casimir Effect}
Following the standard recipe \cite{schwinger2}, we evaluate the Casimir energy density   $\langle \epsilon_{Cas}\rangle =-\lim_{\nu\rightarrow 0}\frac{1}{AL}\frac{\partial}{\partial\tau} \Re {\rm e}\, W(\nu)$. After a $\beta$-power expansion, we get \cite{grad,nist}
\be
\langle \epsilon_{Cas}\rangle =-\frac{\pi^{3/2}}{16L^4}\sum_{k=0}^\infty \frac{2^k\xi^{2k}a_k}{(1-\xi\tau)^{2k}}\bigg(\frac{L}{\pi}\bigg)^{2k}\Gamma\big(-\frac{3}{2}+k\big)\zeta(-3+2k).\label{statcas}
\ee
Taking the leading ($k=0$) and the next to leading order ($k=1$) term, we have
\be\label{mainresult}
\langle \epsilon_{Cas}\rangle=-\frac{\pi^2}{1440L^4}+\frac{1}{384L^2}\frac{\xi^2}{(1-\xi\tau)^2}+O(\xi^4).
\ee
The first term is the (expected) usual flat Casimir energy density. At the horizon crossing ($\tau\rightarrow 0^-$), we have [recall that $\xi=3/(2r_g)$]
\be
\langle \epsilon_{Cas}\rangle_{hor}=-\frac{\pi^2}{1440L^4}\bigg[1-\frac{135}{(4\pi)^2}\bigg(\frac{L}{r_g}\bigg)^2\bigg].
\ee
Eq.(\ref{mainresult}) tells us how the corrections to the Casimir energy density change with the proper time as the cavity approaches the black hole horizon. The above result shows that the comoving observer measures a small reduction in the (absolute) value of the (negative) Casimir energy near the black hole horizon. At a first glance, this may seem rather puzzling, as one would expect no change with respect to the usual flat spacetime result $\langle\epsilon_{Cas}\rangle_{stat}=-\frac{\pi^2}{1440L^4}$ for a  {\em freely falling} Casimir cavity, due to the Equivalence Principle. We will comment about such issue in the last section.

\section{Bunch-Davies Vacuum and Particle Creation}
Particle creation is related to the vacuum persistence amplitude, i.e., the imaginary part of the effective action $W$. In the in-out formalism we have
\be
|\langle {0\, \rm out}|{0\, \rm in}\rangle|^2=e^{2i\Im {\rm m}W},
\ee
so that the (usually small) number density of created particles inside the falling cavity is $\langle n\rangle\simeq \frac{2\,\Im {\rm m}\, W}{AL}$.
Expanding the imaginary part of $W(\nu)$ we obtain
\be
\Im {\rm m}W=\frac{A}{24\pi^3 L^2}\bigg[-\frac{\xi^2 L^2}{(1-\xi\tau)^2}\zeta (1)+\frac{2\xi^4 L^4}{15(1-\xi\tau)^4}+\cdots\bigg].
\label{WGfinal}
\ee
Inspection of (\ref{WGfinal}) reveals that the first  term in the square brackets is {\em divergent}. 

In spite of the divergent term, when the small adimensional quantity $\xi L= \frac{3L}{2r_g}$ is vanishing then $\Im {\rm m}\, W\rightarrow 0$, hence implying no particle creation inside the falling cavity, as expected. Actually, when the gravitational radius of the black hole is much greater than the plate separation, the cavity does not experience any relevant effect due to the free fall (the cavity {\em feels} an almost {\em uniform} gravitational field).
We will avoid the difficulties stemming from the appearance of infinities in the imaginary part of the effective action exploiting the relationship between the Schwinger theory and the in-out formalism, based upon the Bogolubov  approach.

Observe that the field modes (\ref{chitot}) have the required minkowskian (plane wave) behaviour at $\tau\rightarrow -\infty$, when the cavity is at the spatial infinity with respect to the black hole.
They satisfy the Bunch-Davies vacuum requirements \cite{parker2,fulling2,davies}, admitting a plane wave solution in the far past
\be\label{chiplane}
\chi_k(\tau)=\frac{\alpha}{\sqrt{2\omega_k}}e^{-i\omega_k\tau}+\frac{\beta}{\sqrt{2\omega_k}}e^{i\omega_k\tau}.
\ee
 Matching (\ref{chitot}) and (\ref{chiplane}) nearby the black hole horizon ($\eta \geq 1$), we easily compute the Bogolubov coefficients $\alpha$ and $\beta$, thus obtaining \footnote{An interesting approach, requiring no detailed knowledge of state normalization,  based upon the paper by Hamilton et al. [A. Hamilton, D. Kabat and M. Parikh, JHEP 0407, 024 (2004)], may be used as well to obtain the same result.}
\be\label{bog}
|\alpha_k|^2=1+\frac{\xi^2}{16\omega_k^2 (1-\xi\tau)^2},\quad\quad\quad |\beta_k|^2=\frac{\xi^2}{16\omega_k^2 (1-\xi\tau)^2},
\ee
satisfying $|\alpha_k|^2-|\beta_k|^2=1$. The $\beta$ coefficient is related to particle creation. Note that, as $\tau\rightarrow -\infty$,  $|\alpha_k|^2\sim 1$ and $|\beta_k|^2\sim 0$, i.e., we have no particle creation in the far past, as expected, meanwhile at the horizon crossing ($\tau =0$)  we have $|\beta_k|^2=\frac{\xi^2}{16\omega_k^2}$. 
Although the number of created particles is a divergent quantity, we can get a {\em finite} result for the energy density $\langle \epsilon_{\rm dyn}\rangle$  of the created quanta, writing 
\be
\langle \epsilon_{\rm dyn}\rangle=\frac{1}{AL}\bigg[\frac{A}{(2\pi)^2}\sum_n\int d^2 k_\perp\frac{\xi^2}{16\omega_k^2\eta^2}\omega_k\bigg]=\frac{\xi^2}{384L^2(1-\xi\tau)^2}.
\label{DCE}
\ee
Comparing the above result with (\ref{mainresult}), describing the vacuum energy density pertaining to the Casimir effect, 
we see, quite interestingly, that the small reduction observed in the static Casimir energy value is {\em just the same} as the amount of energy of created field particles. This could suggest a close relationship between the two considered effects. Nevertheless, some care is required when speculating about such coincidence, as both the results have been obtained up to the first-order approximation \cite{fs}.

\section{Concluding remarks}
We have considered the Casimir energy density corrections in a small cavity freely falling from the spatial infinity into a Schwarzschild black hole. 

At a first glance, one could wonder that corrections to the static Casimir effect as well as particle creation are detected by an observer in a freely falling {\em inertial} frame.  However, as anticipated in the introduction, this is not too surprising. The Equivalence Principle (EP), deeply rooted in the theory of General Relativity (GR), applies well in the context of a {\em local} theory, just as GR is. On the other hand, when quantum fields are taken into account, the {\em non-local} character of of the underlying quantum theory conflicts with the EP, causing the latter to be not straightforwardly applicable \cite{fulling3}. 

In the present scenario, the effective action $W$  behaves as a {\em non-local} object, thus probing the global spacetime structure, through the long-wavelength field modes. The adopted renormalization procedure (whatever it may be) does transfer the  spacetime details into a renormalized stress-energy tensor, which is eventually the locally {\em measured} quantity. In such a way, information contained in the spacetime geometry surrounding the cavity bypasses - so to say -  the EP, appearing both in the form of a small correction to the expected static Casimir energy and a tiny flux of created field quanta.

In deriving the above results several assumptions have been made. In particular, we have neglected other possible contributions related to the cavity extension. Tidal effects, e.g., are expected to give rise to anisotropies in the energy density distribution inside the cavity; such aspect will be  considered in detail (in the case of a 1+1D model) in a forthcoming paper \cite{justin}.

It maybe that some of the divergencies met throughout the paper are due to the (in)finiteness of the Casimir plates. In that respect, it seems likely that a more physically consistent analysis, based upon a  {\em finite} 3D cavity, could lead to remove the residual infinities encountered in the present approach. 

Finally, an obvious improvement of the present research would be to extend the analysis to appear in \cite{justin} to the 3+1D case, including both {\em tidal} and {\em 3D-finite-size} effects in evaluating the corrections to the Casimir effect. This we hope will be our next goal.\\

\end{document}